\documentclass[10pt,conference,final]{IEEEtran}

\usepackage[utf8]{inputenc}
\usepackage{algorithm}
\usepackage[noend]{algpseudocode}
\usepackage{amsmath}
\usepackage{graphicx}
\usepackage{color}
\usepackage{xcolor}
\usepackage{multirow, makecell}
\usepackage{amssymb}
\usepackage{bold-extra}
\usepackage[url=false,doi=false,isbn=false,eprint=false,firstinits=true,maxnames=10]{biblatex}
\DeclareNameAlias{sortname}{last-first}

\newcommand{\ignore}[1]{}

\newcommand{\satish}[1]{{\color{blue} [~\emph{SC:~#1}~]}}
\newcommand{\revision}[1]{{\color{black} ~#1}}

\newcommand{\name}{Bug2Commit}

\newcommand{\locus}{Locus}
\newcommand{\locusq}{Orca}
\newcommand{\nameBM}{\name}
\newcommand{\nameft}{B2C-fasttext}

\newcommand{\mytt}[1]{\footnotesize \texttt{#1}}

\begin{document}
\title{Industry-scale IR-based Bug Localization: A Perspective from Facebook}

\author{
\IEEEauthorblockN{Vijayaraghavan Murali}
\IEEEauthorblockA{Facebook, Inc.\\
U.S.A. \\
vijaymurali@fb.com}
\and
\IEEEauthorblockN{Lee Gross}
\IEEEauthorblockA{Facebook, Inc.\\
U.S.A. \\
leegross@fb.com}
\and
\IEEEauthorblockN{Rebecca Qian}
\IEEEauthorblockA{Facebook, Inc.\\
U.S.A. \\
rebeccaqian@fb.com}
\and
\IEEEauthorblockN{Satish Chandra}
\IEEEauthorblockA{Facebook, Inc.\\
U.S.A. \\
schandra@acm.org}
}

\maketitle

\date{August 2020}

\begin{abstract}

We explore the application of Information Retrieval (IR) based bug localization methods at a large industrial setting, Facebook.
Facebook's code base evolves rapidly, with thousands of code changes being committed to a monolithic repository every day.
When a bug is detected, it is often time-sensitive and imperative to identify the commit causing the bug in order to either revert it or fix it.
This is complicated by the fact that bugs often manifest with complex and unwieldy features, such as stack traces and other metadata.
Code commits also have various features associated with them, ranging from developer comments to test results.
This poses unique challenges to bug localization methods, making it a highly non-trivial operation.

In this paper we lay out several practical concerns for industry-level IR-based bug localization, and propose \name{}, a tool that is designed to address these concerns.
We also assess the effectiveness of existing IR-based localization techniques from the software engineering community, and find that in the presence of complex queries or documents, which are common at Facebook, existing approaches do not perform as well as \name{}.
We evaluate \name{} on three applications at Facebook: client-side crashes from the mobile app, server-side performance regressions, and mobile simulation tests for performance.
We find that \name{} outperforms the accuracy of existing approaches by up to 17\%, leading to reduced time for triaging regressions and attributing bugs found in simulations.

\end{abstract}

\section{Introduction}
\label{sec:intro}

Facebook operates a monolithic repository of code to which developers commit thousands of code changes every day.
Each commit passes through standard industrial processes such as code review and continuous integration.
Despite these sentinels, commits containing bugs invariably make their way into the repository.
In certain cases, such as in server-side code, the bugs can manifest quickly by causing a sudden change in a continuously tracked performance metric.
In other cases, such as client-side code, the bugs can take up to weeks to manifest due to the need for users to update and use the buggy app version.
In either case, once a symptom of the bug is detected, there is a need to identify the point at which the bug was introduced.
This is the problem of {\em bug localization}.

Bug localization has been a long studied problem in software engineering literature, where recent advances can be classified into two types: {\em spectrum-based} methods~\cite{Jones2002, abreu2007accuracy} that work with execution trace data, and {\em information retrieval (IR)-based} methods~\cite{Zhou2012,Saha2013,Lam2015,Bhagwan2018} that work with the textual similarity between bug reports and code.
In this work, we focus on IR-based localization methods as we do not have access to execution traces in our setting.
IR-based methods typically cast bug localization as a search problem, where the search {\em query} consists of terms from the bug report, and the {\em documents} to retrieve are code entities in a repository.
There exists a large body of work~\cite{Zhou2012,Saha2013,Lam2015,Rahman2018,pradel2020scaffle} along this line, on localizing the bug to locations in the code base, using a mix of static analysis or learning-based models.

To deploy any bug localization method in our setting, we first observe that it must have the following desired properties:

{\em 1) The method must localize to commits, not to files.}
In a large industrial setting such as Facebook, bug localization is often a time-sensitive or compute-intensive operation.
In cases where the bug causes a crash or performance regression in production code, an ``on-call engineer'' is responsible for triaging the bug to the right team or developer who is capable of handling the issue.
Triaging it to the wrong developers wastes precious time and resources in getting the issue fixed, and could lead to large service disruptions.
Sometimes the bug is detected in simulation runs of product builds, in which case a bisect operation is triggered to find the right culprit commit in the build.
These builds typically consist of several hundreds of commits, and each bisect step consumes a significant amount of time and compute resources to compile, build, and run the simulation.
In both these cases, the most important operation is to {\em find the commit that introduced the bug}.
The next step would typically be to revert the commit and/or assign the bug to the author of the commit.
Localizing to files does not help in this process, as in a large-scale setting a single file could be ``owned'' by multiple developers and several commits could have made changes to it within a small time frame.

{\em 2) The method must be unsupervised.}
At Facebook, when developers are assigned a bug, they have an incentive to fix the bug as soon as possible in order to minimize service disruptions or maintain product quality.
Due to this, there exists a significant amount of historical data on bug {\em fixes}, which methods like Scaffle~\cite{pradel2020scaffle} can utilize.
However, developers are not required to investigate and mark the culprit commit that originally introduced the bug during their process of bug fixing.
This leads to a scarcity of data on bug {\em causes}.
The SZZ algorithm~\cite{szz,williams2008} was proposed to extract bug inducing commits from fixes by traversing the commit graph, but its effectiveness has been debated~\cite{Neto2018}.
Therefore, due to this lack of labeled training data, supervised learning-based approaches for bug localization are not suitable in our setting.

\ignore{
{\em 3) The method must not rely on static analysis.}
Facebook's code base is large and highly heterogeneous~\cite{pradel2020scaffle}, with millions of lines of code written multiple programming languages residing in the same repository.
As a result, a single commit can comprise of code from multiple languages, be event-driven, depend on cross-language queries, or call into back-end services.
Even performing a lightweight static analysis at this scale and complexity is far from a trivial operation.
Therefore, any localization method we deploy must not rely on static analysis.\satish{This point may be unnecessary. I don't this static analyses are designed to do fault localization anyway.}
}

{\em 3) The method must be able to handle complex queries and documents.}
The localization problem in our setting is compounded by the fact that symptoms of bugs usually contain a variety of {\em features} originating from different sources.
For instance, a single bug report can contain a mixture of multiple stack traces from different threads, along with the names of regressing metrics.
Moreover, code commits also come with various features associated with the code change itself, such as natural language descriptions of the commit, code review comments, and test execution results.
Each feature comes with its own characteristics, for instance, stack traces tend to be long whereas the names of regressing metrics are quite succinct.
This means that a ``query'' or ``document'' in our setting is not a monolithic bag-of-words, but rather a complex entity with multiple features.
In fact, it is quite possible for a query in our setting to be more complex than the document itself.
This requires any localization method we use to be designed to handle these complex entities.

{\em 4) The method must be capable of understanding the semantics of features.}
Facebook's code base is large and highly heterogeneous~\cite{pradel2020scaffle}, with millions of lines of code written in multiple programming languages residing in the same repository.
Coupled with a fast-paced development environment, this leads to idiosyncratic coding conventions being used in various parts of the repository, posing a unique challenge to IR-based methods that work with word similarity.
For instance, in one of our applications, we noticed that the terms appearing in bug reports and commits could seem different but actually share the same meaning, such as ``IG'' and ``Instagram''.
Sometimes, the terms could appear concatenated together, such as ``videoplayerinterface'', making it difficult to match them with their component terms.
Developers might also make typographical errors in natural language features such as the commit summary.
Therefore, any localization method we use should be able to capture the semantics of words appearing in features, going beyond the observable set of terms.

In essence, this multi-modality of features in bug reports and commits, coupled with various sources of noise and the lack of labeled training data, makes it highly challenging for IR-based methods to localize the bug to a culprit commit.
Due to this, many existing methods in the literature are not suited to be applied in our setting, especially in an off-the-shelf manner.
Table~\ref{table:related} shows a summary of the characteristics of existing work vis-\`a-vis our desired properties.

\begin{table}
    \centering
    \begin{tabular}{|p{1.5cm}|p{1.5cm}|p{2cm}|p{2cm}|}
    \hline \hline
    \multicolumn{3}{|c|}{\bf Characteristics} & \multicolumn{1}{c|}{\bf Work} \\
    \hline \hline
    \multicolumn{3}{|l|}{Localize to lines (spectrum based)} &
    \cite{Jones2002, abreu2007accuracy} \\ \hline
    \multicolumn{3}{|l|}{Localize to files (IR based)} & \cite{Nguyen2011,Zhou2012,Saha2013,Wu2014,Ye2014,wong2014boosting,Moreno2014,Lam2015,Rahman2018,Koyuncu2019}
    \\ \hline
    Localize to commits (IR based) & \multicolumn{2}{l|}{Supervised learning-based} & \cite{Wu2018,Bhagwan2018} \\ \cline{2-4}
    & Unsupervised & Simple features, No semantics & \cite{Wen2016} and tf-idf model from \cite{Bhagwan2018} \\ \cline{3-4}
    & & Complex features and semantics & \textbf{This work} \\
    \hline \hline
    \end{tabular}
    \caption{Characteristics of existing bug localization methods.}
    \label{table:related}
    \vspace{-5mm}
\end{table}

For instance, most of the prior work using IR-based methods have been applied to localize bugs to files~\cite{Nguyen2011,Zhou2012,Saha2013,Wu2014,Ye2014,wong2014boosting,Moreno2014,Lam2015,Rahman2018,Koyuncu2019}.
Localizing bugs to code commits instead, a concept termed {\em commit-level} bug localization, has been fairly recent~\cite{Wen2016,Wu2018,Bhagwan2018}.
Among this line of work, ChangeLocator~\cite{Wu2018} uses a supervised learning-based approach, not satisfying our requirements.
Orca~\cite{Bhagwan2018} also uses a learning approach to break ties in ranking, but its main tf-idf model, and Locus~\cite{Wen2016}, are more aligned with our requirements, and so we chose to evaluate their effectiveness.
However, we found that they are not adequate in our setting, as they are not efficient at either handling complex entities or understanding semantics.
This raises the need for us to design a bug localization method specifically with these desired properties in mind.

Towards this goal, we make the following contributions:

\begin{itemize}
    \item We propose \name{}, an unsupervised commit-level bug localization method that builds upon IR-based techniques but is adept at handling complex queries and documents, and at understanding the semantics of features.
    
    \item We discuss several technical design decisions that went into \name{}.
    We look at vector space models (VSMs), multiple choices of vectorizers to use, and vectorizing bug reports and code commits in the presence of complex entities.
    
    \item We evaluate \name{} on three applications at Facebook: (i) client-side crashes from Facebook's mobile apps, (ii) server-side performance regressions, and (iii) performance simulation tests on mobile code.
    We find that \name{} can outperform existing methods in accurately identifying the culprit commit by up to 17\%.
    
    \item We show that \name{} leads to more optimized development operations at Facebook: regression triage time for oncall engineers is reduced from hours to minutes, and the run time of bisect operations to find the culprit commit is reduced significantly.
\end{itemize}

\ignore{
The rest of the paper is organized as follows.
Section~\ref{sec:overview} sets the problem context of bug localization at Facebook with an example.
Section~\ref{sec:technical} presents technical details of the \name{} model that addresses the aforementioned practical considerations.
Section~\ref{sec:implementation} discusses details of the implementation of \name{}.
Section~\ref{sec:evaluation} presents results from \name{}'s evaluation at Facebook.
Section~\ref{sec:related} discusses related work.
Finally, Section~\ref{sec:conclusion} concludes the paper.
}

\section{Problem Context \& Example}
\label{sec:overview}

\begin{figure*}
    \centering
    \includegraphics[width=1.6\columnwidth]{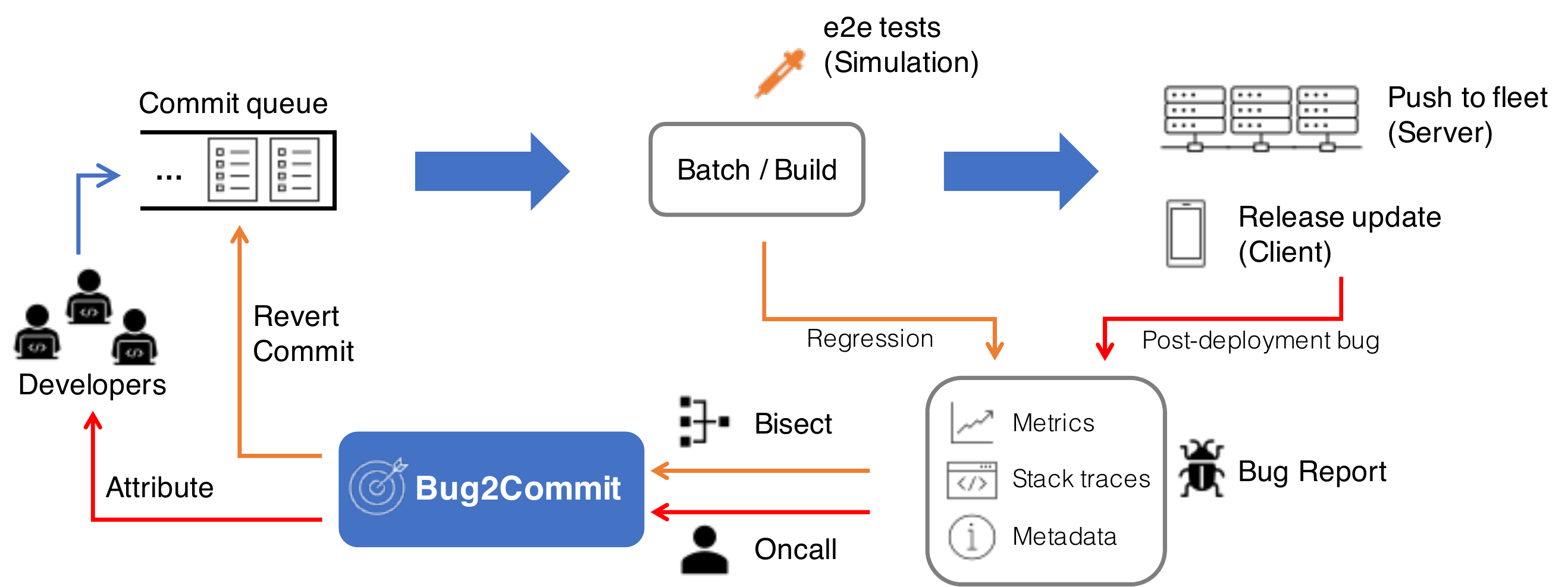}
    \caption{Software deployment at Facebook and bug localization with \name{}}
    \label{fig:overview}
\end{figure*}

\begin{table*}
    \centering
    \begin{tabular}{c}
    \begin{tabular}{|c|c|l|}
    \hline \hline
    {\bf Content Type} & {\bf Source} & \multicolumn{1}{c|}{\bf Description} \\
    \hline \hline
    Stack trace & Crash & The stack trace at the time of crash, typically reported from the client side mobile OS. \\ \hline
    Exception & Crash & Exception message and other information associated with the stack trace. \\ \hline
    Metadata & Crash & Information about the environment in which the crash occurred, e.g., app surface the user was in. \\ \hline
    Metric & Regression & Name of the metric regressing in performance, e.g., the name of a regressing function or performance test. \\ \hline
    Regressing traces & Regression & Snapshots of traces from machines where the regression was detected. \\ \hline
    Task description & Crash/Regression & If a task has been created to track the bug, typically automatically, its description. \\ 
    \hline \hline
    \end{tabular}
    \\ (a) \\
    \begin{tabular}{|c|c|l|}
    \hline \hline
    {\bf Content Type} & {\bf Source} & \multicolumn{1}{c|}{\bf Description} \\
    \hline \hline
    Title \& summary & Code commit & The developer-provided (natural language) title and summary of a code commit. \\ \hline
    Test plan & Code commit & Commands provided by the developer to validate the commit's changes. \\ \hline
    Filenames & Code commit & Names of files, typically translating to classes, changed in the commit. \\ \hline
    Config name & Config change & Name of the server-side configuration that has been changed. \\ \hline
    References & Config change & Snapshots of stack traces of functions that reference the configuration. \\ \hline
    \hline \hline
    \end{tabular}
    \\ (b) \\
    \end{tabular}
    \caption{(a) Features appearing in bug reports, and (b) features appearing in commits.}
    \label{table:features}
    \vspace{-7mm}
\end{table*}

In this section, we set the problem context with an overview of bug localization at Facebook and an example using \name{}.

\subsection{Problem Context}
\label{subsec:overview}

\subsubsection{Software Deployment at Facebook}
\label{subsubsec:sdlc}

Fig.~\ref{fig:overview} shows Facebook's software development life-cycle.
Thousands of developers commit code that goes through the cycle of code review, unit testing, and static analysis.
Once the developer and reviewer are satisfied with the commit, it is then {\em landed} by the developer, at which point it is handed over to continuous integration.
This system maintains a queue of commits that need to be merged onto the monorepo.
At regular intervals, a particular batch of commits is selected in order to be subjected to continuous performance testing.
These computationally expensive tests replay simulations of user interactions on the app surface~\cite{mobilelab}.

Once the tests are complete, the batch of commits is then merged onto the master branch.
Server-side commits are pushed on to Facebook's fleet of machines, whereas 
client-side code is packaged into a release and sent to mobile app stores.

\subsubsection{Bug Localization at Facebook}
\label{subsubsec:localization}

Bugs are typically detected at two stages of software deployment.
First, the simulation tests can catch bugs before a batch of commits is deployed.
During this testing process, a number of app and server health metrics are monitored for changes.
Any anomalous movement of a metric is flagged and raised as a regression to be investigated.
Secondly, since these tests do not completely exercise all behaviors, post-deployment bugs are inevitable.
On the server side these bugs manifest as performance regressions, and can typically be detected in a few hours as the code would be exercised by queries from millions of users.
On the client side it could take several weeks for post-deployment bugs to surface due to the longer app release cycle, and users needing to update to the latest version.

Whether it is a regression or post-deployment bug, a {\em bug report} is generated.
The format of these bug reports can vary depending on the actual nature of the bug. 
At a high level, bug reports contain various pieces of information from different sources.
They can contain metrics that were found to regress, snapshots of stack traces at the time of occurrence of the bug, and metadata about the environment in which the bug occurred.
As these bug reports are automatically generated, they do not typically contain pithy natural language descriptions of the bug.
Table~\ref{table:features}(a) provides a list of commonly occurring features in bug reports.

A bug report is acted upon in two possible ways, as portrayed in Fig.~\ref{fig:overview}.
If it is a regression detected during the simulation process, a bisect operation is triggered on the whole build to find the right culprit commit.
This commit would then be reverted from the build and the author of the commit would be notified.
If the bug is from post-deployment, an on-call engineer is assigned the bug report, whose goal is then to attribute the bug to the correct developer or team for putting out a fix.

\subsubsection{Role of \name{}}
\label{subsubsec:c2d-role}

\name{} is invoked in the pipeline by either the automated bisect system or an oncall engineer.
Before invoking \name{}, a set of {\em candidates} are gathered, consisting of commits among which the bug-inducing commit is present.
The candidates are collected using the time of the regression, or the build in which a bug is first detected.
The typical size of this candidate set could range from a few hundred to thousands.
Candidates can either be code commits or configuration (``config'') changes.
Code commits make changes to the code base as in any version control system, whereas config changes make changes to key-value style configuration of systems which are referenced throughout the code base\footnote{For simplicity, we refer to code changes and config changes as ``commits''.}.
Candidates also come with a complex set of features, as shown in Table~\ref{table:features}(b).
The goal of \name{} is then as follows: given a bug report and a candidate set of commits, accurately identify the culprit commit that introduced the bug.


\subsection{Example}
\label{subsec:example}

\begin{figure*}
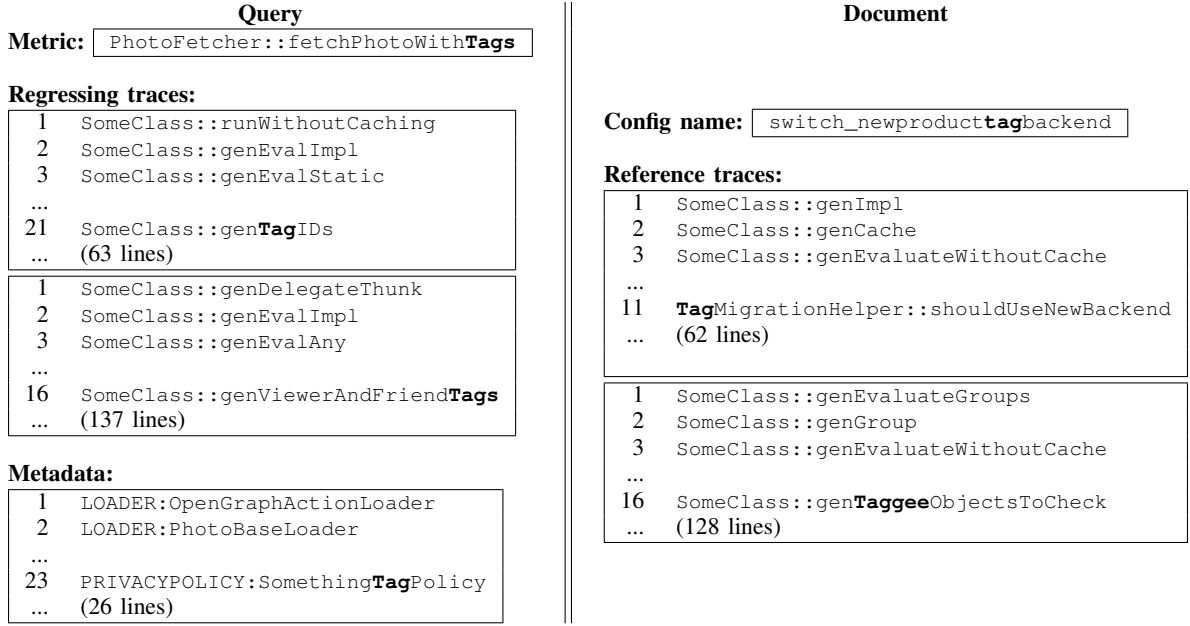

\centering
\small
\begin{tabular}{l||l}

\multicolumn{1}{c||}{\bf Query} & \multicolumn{1}{c}{\bf Document} \\

\begin{tabular}{l}

{\bf Metric:}

\begin{tabular}{|c|} \hline
\mytt{PhotoFetcher::fetchPhotoWith\textbf{Tags}} \\ \hline
\end{tabular}

\\ \\

{\bf Regressing traces:} \\

\begin{tabular}{|rl|} \hline
1 & \mytt{SomeClass::runWithoutCaching} \\
2 & \mytt{SomeClass::genEvalImpl} \\
3 & \mytt{SomeClass::genEvalStatic} \\
... &  \\
21 & \mytt{SomeClass::gen\textbf{Tag}IDs} \\
... & (63 lines)

\\ \hline \hline

1 & \mytt{SomeClass::genDelegateThunk} \\
2 & \mytt{SomeClass::genEvalImpl} \\
3 & \mytt{SomeClass::genEvalAny} \\
... &  \\
16 & \mytt{SomeClass::genViewerAndFriend\textbf{Tags}} \\
... & (137 lines) \\ \hline
\end{tabular}

\\ \\

{\bf Metadata:} \\

\begin{tabular}{|rl|} \hline
1 & \mytt{LOADER:OpenGraphActionLoader} \\
2 & \mytt{LOADER:PhotoBaseLoader} \\
... &  \\
23 & \mytt{PRIVACYPOLICY:Something\textbf{Tag}Policy} \\
... & (26 lines) \\ \hline
\end{tabular}

\end{tabular}

& 

\begin{tabular}{l}

{\bf Config name:}

\begin{tabular}{|c|} \hline
\mytt{switch\_newproduct\textbf{tag}backend} \\ \hline
\end{tabular}

\\ \\

{\bf Reference traces:} \\

\begin{tabular}{|rl|} \hline
1 & \mytt{SomeClass::genImpl} \\
2 & \mytt{SomeClass::genCache} \\
3 & \mytt{SomeClass::genEvaluateWithoutCache} \\
... &  \\
11 & \mytt{\textbf{Tag}MigrationHelper::shouldUseNewBackend} \\
... & (62 lines) \\

& \\ \hline \hline

1 & \mytt{SomeClass::genEvaluateGroups} \\
2 & \mytt{SomeClass::genGroup} \\
3 & \mytt{SomeClass::genEvaluateWithoutCache} \\
... &  \\
16 & \mytt{SomeClass::gen\textbf{Taggee}ObjectsToCheck} \\
... & (128 lines) \\ \hline
\end{tabular}

\end{tabular}

\end{tabular}
\caption{Example of a query (bug report) and a document (candidate commit) with complex features.}
\label{fig:example}
\end{figure*}

Fig.~\ref{fig:example} shows an example of a bug report arising from a server-side performance regression, and the configuration change that had caused it.
Notice that the query in our setting is far more complex than a succinct set of keywords or a single crash stack trace, as assumed by many IR-based methods.
For example, if only the metric name is used as a succinct query, it might indicate importance of the word ``photo'' or ``fetch'' as they occur prominently in the metric.
However, this would have been misleading, as utilizing other features in the bug report reveals that the word ``tag'' is actually a more important and relevant term for the query (highlighted for readability).

One naive way to include all features in a query is to treat them together as a single monolithic bag-of-words.
However, doing this would again dilute the importance of ``tag'' as words such as ``eval'' or ``loader'' would appear to carry more importance.
To address this problem, \name{} uses a technique that vectorizes each feature individually, and then aggregates the resulting vectors.
Also of note are the different characteristics of each feature: the metric name is pithy and high signal, whereas the regressing traces are long and noisy, and the metadata is somewhere in between.
One of the goals of \name{} is, therefore, to balance bringing in more information from long and noisy features, while still not diluting the signal from short and succinct features.
To do this, \name{} uses a BM25-based~\cite{Robertson96okapi} method that is designed to consider the size of each feature when computing the importance of terms.

In addition to multiple features, our practical setting imposes more difficulties while attempting to match terms between the query and candidate documents.
As seen in Fig.~\ref{fig:example}, the word ``tag'' appears in a synonymous form, as ``taggee'', in the culprit commit.
IR-based methods typically pre-process their input through a stemmer and/or lemmatizer, but these are limited in their ability to understand the semantics of words.
For example, the Porter stemmer used in~\cite{Wen2016} would not be able to stem ``taggee'' to ``tag'', which would actually require understanding that they are synonymous in nature.
To handle this, \name{} uses weighted word embeddings to represent words with a cognizance of their semantics.
Finally, the config change also has the word ``tag'' appearing in an obfuscated form, such as in the concatenated name ``newproducttagbackend''.
Handling this would require a smarter pre-processing step, which we will come to later in Section~\ref{sec:implementation}.

\section{Technical Details}
\label{sec:technical}

In this section, we set up technical preliminaries about existing IR-based methods, and present \name{}'s approach that addresses practical concerns with existing methods.

\subsection{Background on Vector Space Models (VSMs)}
\label{subsec:vsm}

Existing IR-based methods localize bugs by relying on the textual similarity of terms appearing in bug reports and commits, the idea being the more overlap there is the more relevant a commit is to a bug.
Several IR-based methods have found success with {\em vector space models (VSMs)}~\cite{Wen2016,Saha2013,Zhou2012}.
VSM-based approaches start with defining a {\em vector space} $\mathcal{V}$ based on the vocabulary of all words appearing in the corpus of bug reports and commits.
Given a set of candidate commits $C = \{c_1, \ldots, c_n\}$, each $c_i$ is transformed, or {\em vectorized}, into a sparse vector $v_{c_i} = {\sf vectorize}(c_i)$ in $\mathcal{V}$.
Given a bug report $b$, it is similarly transformed into another vector $v_b = {\sf vectorize}(b)$.
Then, the similarity between the bug report and a candidate is defined using standard metrics such as {\em cosine similarity}:
$
{\sf sim}(v_b, v_{c_i}) = \frac{v_b . v_{c_i}}{||v_b||~||v_{c_i}||}
$

The important step in this process is the ${\sf vectorize}$ operation.
Existing approaches typically use the {\em term frequency (tf)} and {\em inverse document frequency (idf)} of words appearing in the corpus.
Suppose that $V$ is the vocabulary of all words in the corpus of both commits and bug reports.
The vectorization of a commit $c$ is done as follows:
\begin{align} \begin{split}
\label{eqn:tfidf}
{\sf tf}(w, c) = 1 + \log {\sf freq}(w, c) ~~~~~ {\sf idf}(w) = \log \frac{N}{{\sf num}(w)} \\
{\sf vectorize}(c) = \langle {\sf tf}(w, c) \times {\sf idf}(w) ~|~ w \in V \rangle
\end{split}
\end{align}
where ${\sf freq}(w, c)$ is the number of occurrences of word $w$ in commit $c$, $N$ is the total number of commits, and ${\sf num}(w)$ is the number of commits that contain word $w$.
The vectorization of a bug report $b$ is done analogously.
The intuition behind this tf-idf technique is that the more frequent a word is in a document, the more importance it carries, and the more frequent a word appears throughout the corpus of documents, the less importance it carries.
tf-idf allows these two quantities to interplay in computing the weight of words in a document.

\subsection{\name{}: Addressing Practical Considerations}
\label{subsec:practical}

\name{} builds upon VSM-based approaches described above by being designed to address important considerations that arise in our industrial setting.

\subsubsection{Handling complex queries and documents}

In our setting the primary issue with the approach described so far is that a commit or bug report is not a single monolithic entity.
As seen in Table~\ref{table:features} they are composed of various complex {\em features}.
A commit would realistically be represented as $c = \{f_1^c, \ldots, f_n^c\}$ and a bug report as $b = \{f_1^b, \ldots, f_m^b\}$, where each $f_i$ is the value of a feature.
Some features are simple, such as the name of the regressing metric in a bug reports.
Some features are lengthy and complex, such as snapshots of regressing traces.

Treating a bug report as a query or a commit as a document in a monolithic manner would mean coalescing words appearing in all features.
This would make words in lengthy and complex features drown out the weight of words in short but succinct features during vectorization due to having a higher ${\sf tf}$.
This would in turn throw off the similarity metric, contributing to inaccuracy of the model.
In Section~\ref{sec:evaluation} we assess the performance of such a model, and show that it in fact does not perform well.

To overcome this problem, we make two modifications to the vectorization process.
First, we use a special vectorization operation for handling our complex type of documents with features.
Given a commit $c = \{f_1^c, \ldots, f_n^c\}$ (analogously, bug report $b$), we vectorize it as follows:
\begin{align}
\label{eqn:vect}
    {\sf vectorize\_complex}(c) = \frac{1}{n} \sum_{i=1}^n {\sf vectorize}(f_i^c)
\end{align}
Essentially, we vectorize each feature separately and compute the mean of all the resulting vectors.
This results in a vector formed by taking the mean of each co-ordinate across all feature vectors.
In doing so we avoid the cumulative effect of words appearing in multiple lengthy features dominating other words simply through their ${\sf tf}$.

However, this averaging introduces a secondary problem where the weight of important words appearing in simple and succinct features also gets lowered.
To address this, we need a vectorizer that proportionally scales up the weight of a word if it appears in a short feature as opposed to a lengthy feature.
For this purpose, we use a {\em BM25}~\cite{Robertson96okapi} based vectorizer rather than tf-idf.
BM25 is a popular scoring function used by search engines such as Lucene~\cite{luceneBM25}, and has been designed to handle documents with varying length when computing word weights.
Essentially, BM25 tweaks the ${\sf tf}$ and ${\sf idf}$ formulas in Equation~\ref{eqn:tfidf} to take into account the document length~\cite{luceneBM25}.

\ignore{
Given a feature $f$, we vectorize it as follows instead of using Equation~\ref{eqn:tfidf}:
\begin{align} \begin{split}
\label{eqn:bm25}
{\sf tf}(w, f) = \frac{{\sf freq}(w, f) \times (k1 + 1)}{{\sf freq}(w, f) + k1 \times (1 - b + b * \frac{{\sf len}(f)}{L})} \\
{\sf idf}(w) = \log ~ (1 + \frac{N - {\sf num}(w) + 0.5}{{\sf num}(w) + 0.5} )\\
{\sf vectorize}(f) = \langle {\sf tf}(w, f) \times {\sf idf}(w) ~|~ w \in V \rangle
\end{split}
\end{align}
where $k1$ and $b$ are hyper-parameters of BM25, ${\sf len}$ is the length of feature $f$, and $L$ is the average length of features in the corpus.
$k1$ dampens the effect of term frequency, which was in effect achieved by the $\log$ function in Equation~\ref{eqn:tfidf}.
$b$ controls the scaling of a word's weight by considering the length of the feature it appears in, in relation to the average length of features.
If a word appears in a short document, the ratio $\frac{{\sf len}(f)}{L}$ is smaller, making the ${\sf tf}$ value higher.

Together, Equations~\ref{eqn:vect} and~\ref{eqn:bm25} have the effect of weighing important words from short features high while mitigating the cumulative effect of words from lengthy and complex features.
}

\subsubsection{Semantic Embedding of Words}
\label{subsubsec:vectorizer}

VSMs in general treat individual words as unique atomic entities in a vocabulary.
In practice, however, there arise instances where two words look different but share the same semantic meaning.
For instance, developers at Facebook sometimes use ``IG'' as an acronym for ``Instagram''. 
In Section~\ref{sec:evaluation} we will take a closer look at one application where this problem is particularly evident.
The above model would not be able to make this connection, and would treat them as distinct words.
Hence, it would never be able to compare a bug report and commit based on these words, whereas a human would easily be able to.

To address this problem, we use {\em weighted word embeddings}~\cite{sachdev2018retrieval} to vectorize bug reports and commits.
In this method, an embedding function $\phi$ is first trained to map each word in the vocabulary $V$ to a vector in a dense vector space of dimensionality $d$, i.e., $\phi : V \rightarrow \mathbb{R}^d$.
Then, given a feature $f$ (from a commit or bug report), we vectorize it as follows:
\begin{align} \begin{split}
\label{eqn:fasttext}
{\sf vectorize}(f) = \sum_{w \in f} \phi(w) \times ({\sf tf}(w, f) \times {\sf idf}(w))
\end{split}
\end{align}
where ${\sf tf}$ and ${\sf idf}$ are as defined by the BM25 formula.
Essentially, we scale the vector obtained from $\phi(w)$ for word $w$ by its BM25 score.
This ensures that the model still gains the benefits of BM25-based weighting.

The important operation here is to learn $\phi$ in such a way that it maps ``synonymous'' words to nearby vectors in $\mathbb{R}^d$.
This is accomplished using language models, such as the widely used natural language processing (NLP) technique word2vec~\cite{Mikolov2013a}.
We refer the reader to the word2vec paper~\cite{Mikolov2013a} for technical details, but provide a brief overview here.
A language model is trained by making it predict a particular word $w_i$ in a sentence $w_1 w_2 \ldots w_n$ using the context around it.
The context is defined by a window of words around the target word, $w_{i-s} \ldots w_{i-1} w_{i+1} \ldots w_{i+s}$, where $s$ is the window size.
In other words, a language model is trained to predict $\text{Pr}(w_i ~|~ w_{i-s}, \ldots, w_{i-1}, w_{i+1}, \ldots, w_{i+s})$.
To learn $\phi$, each contextual word is represented using its $\phi$-based embedding, which then becomes solving the optimization problem:
$$
\arg \max_\phi \text{Pr}(w_i ~|~ \phi(w_{i-s}), \ldots, \phi(w_{i-1}), \phi(w_{i+1}), \ldots, \phi(w_{i+s}))
$$
As in~\cite{Mikolov2013a}, this objective function is implemented using a neural network, which is then trained on sequences of words appearing in our features.
In Section~\ref{sec:evaluation}, we evaluate the word embeddings based vectorizer for an application where understanding word semantics is important, and show that it outperforms other approaches.

\begin{figure}
    \centering
    \includegraphics[width=\columnwidth]{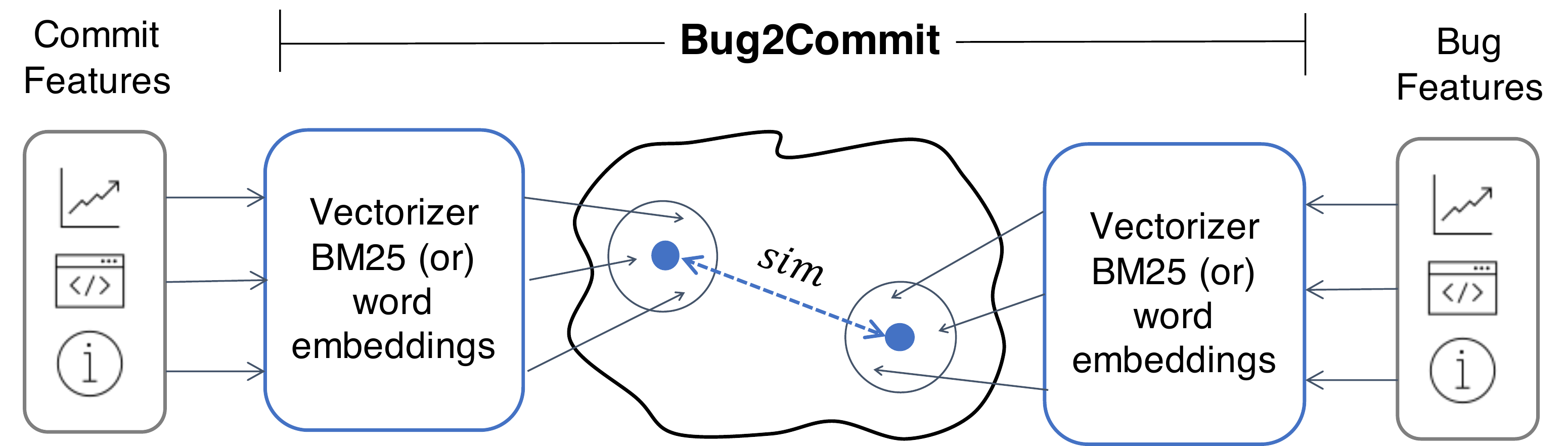}
    \caption{Vectorization of individual features by \name{}}
    \label{fig:c2d-vect}
    \vspace{-2mm}
\end{figure}

In summary, Fig.~\ref{fig:c2d-vect} shows \name{}'s vectorization and localization technique that we discussed in this section.

\section{Implementation}
\label{sec:implementation}

We implemented the model described in Section~\ref{sec:technical} in Python, using the scikit library for vector operations.
Although the model is unsupervised, i.e., it does not rely on labeled data, the implementation has a ``training'' and ``inference'' phase for performance reasons.
In the training phase it continuously gathers a global pool of commits and crashes, and trains word weights for the vectorizer.
This is implemented as a backend pipeline running at a regular cadence.
In the inference phase, \name{} is given a particular bug report to localize, and a  candidate pool of commits among which the culprit commit is present.
This is implemented as a service, suitable to be queried by on-call engineers or systems trying to localizing bugs.
The service keeps the latest trained model loaded up, and is quick to respond to queries, often in a minute or two.

Given a feature, \name{} pre-processes it to extract words from it for vectorization.
While typically a simple step executed using regular expressions that split words by punctuation, CamelCase or snake\_case, this operation becomes tricky in our setting.
As seen in Fig.~\ref{fig:example}, Facebook developers often use concatenated words such as \mbox{``newproducttagbackend''}, sometimes by coding convention in certain parts of the repository, that are not amenable to usual parsing techniques.
%
To address this problem, we use a {\em probabilistic parser} to split the words into its most likely components.
We first compute the probability of seeing a word $w_i$ as $\text{Pr}(w_i) = {\sf count}(w_i) / \sum_{w \in V} {\sf count}(w)$, where ${\sf count}(w)$ is the total number of occurrences of $w$ among all commits and bug reports.
Then, given a word $w$, we infer the most likely split of $w$ as the one that maximizes the product of the probability of the resulting words.
That is, if ${\sf split}(w)$ results in words $w_1, \ldots, w_n$, then we split $w$ according to $\arg \max_{{\sf split}} \prod_{w_i \in {\sf split}(w)} \text{Pr}(w_i)$.
This can be implemented straightforwardly using a dynamic programming algorithm, and so we omit low-level details.

\begin{figure}
    \centering
    \includegraphics[width=\columnwidth]{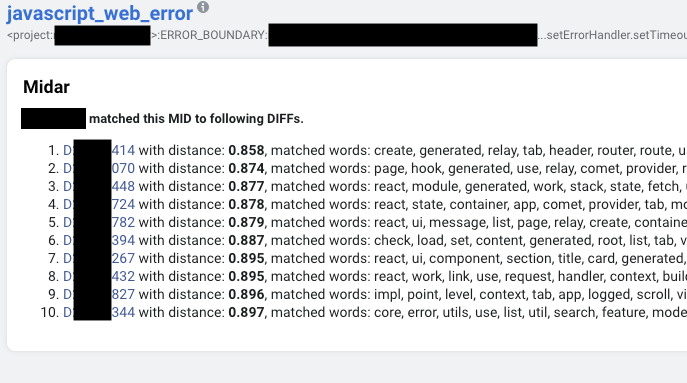}
    \caption{UI for \name{} displaying relevant code commit for a bug.}
    \label{fig:c2d_ui}
    \vspace{-2mm}
\end{figure}

We have developed a user interface for \name{} where the ranked list of relevant commits is displayed, as shown in Fig.~\ref{fig:c2d_ui}.
As it can be seen, in addition to just the list of commits, it also shows the cosine distance (i.e., $1 - $similarity), as a measure of how ``confident'' the model is with the ranking of each result.
Moreover, it also provides an interpretable ``explanation'' of each result using words that contributed the most towards the similarity computation.
On-call engineers have often found this useful to understand why \name{} ranked a particular commit high, and to provide more information to the developer the bug is attributed to.

\section{Evaluation}
\label{sec:evaluation}

\begin{table*}
\centering
    \begin{tabular}{|c||c|c|c||c|c|c|}
    \hline \hline
    {\bf Application} & {\bf Training data} & {\bf Testing data} & {\bf Candidate} & {\bf Training} & {\bf Inference} & {\bf Vectorizer} \\ 
    &&& {\bf pool size} & {\bf time} & {\bf time} & {\bf size} \\ \hline \hline
    
    Mobile app crashes & 4,000 crashes, 25,000 commits & 400 crashes & 1000 commits & 10 mins & $<$ 30 secs & 8,000 words \\
    
    Server perf. regressions & 175 perf. regressions, 11,000 commits & 40 perf. regressions & 300 commits & 5 mins & $<$ 2 mins & 2,800 words \\
    
    Mobile simulation tests & 2250 regressing tests, 150,000 commits & 550 regressing tests & 275 commits & 30 mins & $<$ 3 mins & 35,000 words \\
    
    \hline \hline
    \end{tabular}
    \caption{Characteristics of the data sets for the three applications.}
    \label{table:data}
\end{table*}

In this section, we present results from our evaluation of \name{} at Facebook.
In performing the evaluation, we seek to answer the following research questions:
\begin{itemize}
    \item[RQ1.] How accurate is \name{} in identifying the culprit commits that were the cause of bugs? How does it compare with existing IR-based methods?
    \item[RQ2.] What value does \name{} add to developer operations in a large industrial setting like Facebook?
    \item[RQ3.] In what way do semantic word embeddings contribute to the model?
\end{itemize}

To perform the evaluation, we have collected bug and crash reports at Facebook from three different applications: (i) crashes from the Facebook Android mobile app, (ii) server-side performance regressions, and (iii) mobile simulations tests for performance.
The characteristics of the three data sets are shown in Table~\ref{table:data}.
For each application, using historical data gathered over a period of a year, we perform a train-test split at a particular chronological point in time.
The training and testing data consist of bug reports from crashes or regressions, and commits from the respective candidate pools of those bug reports.
As we will see soon, for each application we evaluate and compare several models.
The training time, inference time, and vectorizer size in Table~\ref{table:data} are shown as an average computed over these models, as there is no significant difference among them.

\subsection{RQ1: Accuracy of \name{}}
\label{subsec:eval-accuracy}

\begin{table}
    \begin{tabular}{|p{0.55in}|l|r|r|r|r|}
    \hline \hline
    {\bf Application} & {\bf Method} & {\bf Top@1} & {\bf Top@5} & {\bf Top@10} & {\bf MRR} \\ \hline \hline
    \multirow{3}{0.55in}{Mobile app crashes}
    & \locus & 0.77 & 0.95 & 0.97 & 0.85 \\
    & \locusq & 0.82 & 0.95 & 0.97 & 0.88 \\
    & \nameBM & {\bf 0.92} & {\bf 0.99} & {\bf 0.99} & {\bf 0.96} \\
    \hline
    \multirow{3}{0.55in}{Server perf. regressions}
    & \locus & 0.63 & 0.71 & 0.83 & 0.69 \\
    & \locusq & 0.63 & 0.71 & 0.83 & 0.69 \\
    & \nameBM & {\bf 0.78} & {\bf 0.88} & {\bf 0.90} & {\bf 0.83} \\
    \hline
    \multirow{3}{0.55in}{Mobile simulation tests}
    & \locus & 0.24 & {\bf 0.46} & 0.53 & 0.35 \\
    & \locusq & {\bf 0.25} & {\bf 0.46} & 0.53 & {\bf 0.36} \\
    & \nameBM & {\bf 0.25} & {\bf 0.46} & {\bf 0.57} & {\bf 0.36} \\
    \hline \hline
    \end{tabular}
    \caption{Evaluation of different methods on top-k accuracy}
    \label{table:results}
\end{table}

In this experiment, we evaluate the accuracy of \name{} in correctly identifying the culprit commit given a bug report and a candidate set of commits.
We use two standard metrics for this:

(i) {\em Top@k}: This measures if the correct culprit commit is present in the top-k results returned.
If the approach has a high top-1 accuracy, the result can often be used for downstream purposes automatically without a human involved.
The workflows of some teams are suitable to adapt to this method, whereas other teams would still want to vet the results through an on-call engineer.
For these cases, recent work on bug localization~\cite{Bhagwan2018} found out that showing top-10 results offers a good balance between result quality and UI succinctness.
Therefore, we choose $k=1,5,10$ for this metric.
    
(ii) {\em Mean Reciprocal Rank (MRR)}: This is defined as the average of the reciprocal rank over the testing data.
If $r$ is the rank at which the correct culprit commit is present in a set of returned results, the reciprocal rank is $1/r$.
The higher this value is, the closer the culprit commit is in general to the top results, leading to less effort required by an on-call engineer for further investigation.

We evaluate \name{} based on these metrics.
In order to also compare with existing IR-based localization methods, we choose the tf-idf based VSM model of Locus~\cite{Wen2016} since it is the closest work related to us.
Orca~\cite{Bhagwan2018} also uses a tf-idf based method but with a minor tweak -- it also ``trains'' the weights using words from the query data, and we also choose this for comparison.
Since the code is not available for these tools, we have re-implemented their models ourselves.

The results are shown in Table~\ref{table:results}.
We find that across all three applications, \nameBM{} performs at least as well or significantly better than both the baselines in both metrics.
The biggest difference we can observe is 17\% (71\% vs 88\%) in the top-5 accuracy in server side performance regressions.
This shows that the fundamental technical difference of \name{} in handling complex entities pays off.
We also observe that for the mobile simulation tests application, \name{} is only marginally better than the other methods, which we will come back to later in this section.

\subsection{RQ2: Value-add of \name{} to industrial processes}

To answer this question, we have conducted an experiment by employing \name{} for developer operations at Facebook and observing its impact for the three applications.

\subsubsection{Mobile app crashes}

In this application, one of the most critical problems faced by on-call engineers is that of a server-side change inducing a client-side regression, which can cause huge spikes in client crashes.
This can happen due to cross-language queries and calls to backend services happening in app code.
Although not common, these sudden client-server spikes are treated as severe events in Facebook, where the entire batch of commits could be halted until the issue is identified and triaged, potentially leading to large-scale disruptions for developers.

When this happens, on-call engineers are fully dedicated to the task of trying to stop the culprit commit from pushing to the entirety of the server fleet.
However, due to differences in the programming languages used, coding conventions, and build processes between server and client code, these regressions are extremely difficult to localize.
The process is largely manual, relying on the on-call engineer's experience, and could take several hours.

Over a time period of 6 months, 7 such server-client regressions were detected, and \name{} helped quickly localize 4 of them to the correct commit.
In 3 out of the 4 cases, the culprit was the top-ranked result, while in the other case it was in the top-10.
Without \name{}, engineers would have spent hours attributing these bugs, whereas with \name{} it was reduced to minutes.
Among the 3 cases that \name{} did not catch, 2 of them were because the symbols in stack trace for the query were corrupted (which was evident in the top words returned by \name{}), and in 1 case \name{} had mis-predicted.

\subsubsection{Server side performance regressions}

In this application, an on-call engineer is tasked to attribute a bug causing a server side performance regression to the correct developers.
This is typically done by assigning the task to the author of the commit that is believed to be causing the bug.
On-call engineers need to manually investigate all possible (hundreds of) candidates, consisting of code commits and config changes, to find the culprit.
This could easily take 1-2 hours for experienced engineers and even longer for newer engineers.

However, with \name{}, 90\% of the time, on-call engineers need to look through just 10 candidates -- a reduction of an order of magnitude -- which can be done in a few minutes.
In addition, if the difference in distance between the top-ranked candidate and the rest is significant, this indicates that \name{} is highly confident that the top result is the culprit.
In such cases, the bug can be auto-triaged to the correct developer, entirely removing the human from the loop.

\subsubsection{Mobile simulation tests}

This application is different from the others in that the consumer of localization is not a human but an automated system.
The goal of this system is to find the correct culprit commit among a batch of commits in which a simulation test regresses in performance, and then attribute the regression to the commit's author as usual.
It employs a parallelized bisect algorithm, splitting the batch into several chunks, and recursively narrowing down the regression to a single commit.
This bisect process takes several simulation steps to find the culprit among hundreds of commits.
Since the tests are long and computationally expensive to run, each simulation step running various tests takes around 3 hours.
This results in a median attribution time of about 9 hours.

\revision{
With \name{}, a list of 10 most likely culprit commits is immediately obtained, and a 3-hour simulation step is triggered on this set in parallel to the original bisect process.
If \name{} correctly caught the culprit, one of the simulations will reveal it, in which case the rest of the simulation steps and bisect are killed and the resources are freed.
Otherwise, the bisect operation proceeds as usual.
Essentially, \name{} uses up additional resources for one simulation step upfront, with a 0.60 probability of avoiding several bisect and simulation steps later.
In practice, this has translated to \name{} reducing the median (mean) attribution time to 3 hours (7 hours) down from 9 hours (18 hours), a reduction of 67\% (61\%), while using the same amount of resources -- a significant benefit at Facebook's scale.
}
\ignore{
With \name{}, a list of 10 most likely culprit commits is immediately obtained, and a 3-hour validation iteration is first triggered on this set prior to the original bisect process.
60\% of the time, \name{} is correct and attribution is complete in just these 3 hours.
40\% of the time, \name{} is incorrect in which case it adds 3 hours to the original process.
Therefore, the expected time to attribute with \name{} is 6.6 hours, down from 9 hours, resulting in a net savings of 27\% along with associated compute resources -- a significant benefit when operating at Facebook's scale.
}

\subsection{RQ3: Benefit of word embeddings}

\begin{figure}
    \centering
    \includegraphics[width=0.8\columnwidth]{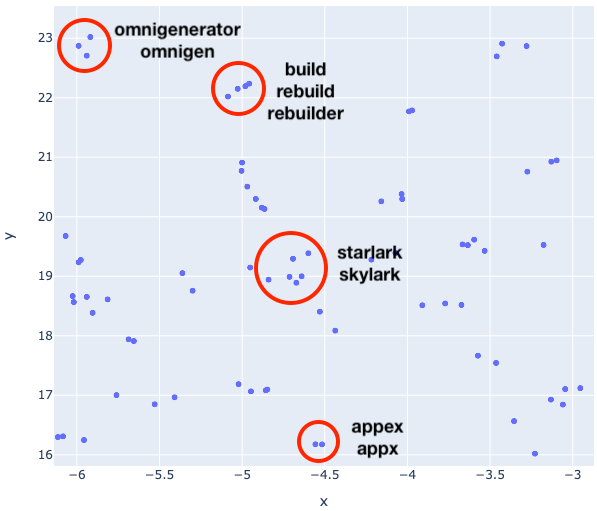}
    \caption{2-dimensional t-SNE plot of semantic word embeddings from fasttext}
    \label{fig:tsne}
    \vspace{-2mm}
\end{figure}

To answer this question we take a closer look at cases where \name{}'s BM25 vectorizer fails to catch the culprit commit in the top-k results.
In mobile crashes and server performance regressions, where \name{} already performed well, these cases primarily involved lack of appropriate signals in the data needed to match terms.
In mobile simulation tests, however, we observed that words did match between the query and candidates, but \name{}, along with other methods, failed to catch them because the words appeared in synonymous forms.
So, for this application, we used the weighted word embedding model of \name{}, implemented using the fasttext~\cite{Bojanowski2017} library.
We used 100 dimensions for the size of the embedding space $d$, i.e., each word in the vocabulary is assigned a particular vector in a 100-dimensional space.
To observe the embeddings visually, we project them onto two dimensions using t-SNE~\cite{maaten2008tsne}.
Fig.~\ref{fig:tsne} plots a section of the resulting space with some groups of words highlighted.

From the plot, we observe that the model captures the semantics of words quite well, and maps synonymous words to nearby points in the space.
For instance, words such as ``omnigen'' and ``omnigenerator'' mean the same to developers, but are simply coming from different parts of the code base with different naming conventions.
Interestingly, the model is also able to capture more complex relationships between words, such as the build library name ``skylark'', which was in fact renamed to ``starlark'' in a later version. 
Finally we also see it capturing typographical errors such as ``appx'' and mapping them to their correct version ``appex''.
This shows evidence that the word embeddings model is learning the semantics of words.
Table~\ref{table:results-fasttext} shows the evaluation of this model (\nameft) in comparison with \name{}'s BM25 vectorizer and other methods.
We see that the model's ability to learn the semantics of words adds a clear benefit in this application, making it outperform all other approaches.

\begin{table}
    \begin{tabular}{|p{0.55in}|l|r|r|r|r|}
    \hline \hline
    {\bf Application} & {\bf Method} & {\bf Top@1} & {\bf Top@5} & {\bf Top@10} & {\bf MRR} \\ \hline \hline
    \multirow{4}{0.55in}{Mobile simulation tests}
    & \locus & 0.24 & 0.46 & 0.53 & 0.35 \\
    & \locusq & 0.25 & 0.46 & 0.53 & 0.36 \\
    & \nameBM & 0.25 & 0.46 & 0.57 & 0.36 \\
    & \nameft & {\bf 0.31} & {\bf 0.50} & {\bf 0.60} & {\bf 0.41} \\
    \hline \hline
    \end{tabular}
    \caption{Accuracy on mobile simluation tests with word embeddings.}
    \label{table:results-fasttext}
    \vspace{-5mm}
\end{table}

\section{Practical Limitations}
\label{sec:limitations}

While \name{} has been successful in its applications at Facebook, the technique does have its limitations in general.
Primarily, since \name{} is based on word (embedding) level overlap between the bug report and candidate commits, it requires a number of features containing relevant words associated with both entities.
Our features in Table~\ref{table:features} were sufficient for our use cases, but in cases where sufficient overlapping words are not present, \name{}'s effectiveness can be greatly reduced.
Secondly, \name{} does not have a good way to weigh words from different features differently.
For instance, it might be useful to provide more weight to words in the title of a bug report than in the hundreds of log lines.
\name{}'s BM25-based vectorizer only uses the frequency of words appearing in the corpus itself to weigh them, but not using domain knowledge such as the above.
Thirdly, as with other machine-learning based approaches, \name{}'s mis-predictions can end up potentially wasting time or resources.
In our applications at Facebook, teams mitigate this effect to some extent by using the confidence of the model in making a prediction (such as in Fig.~\ref{fig:c2d_ui}).
That is, if the model is not highly confident about a prediction, or if the most significant words do not make sense, the oncall would quickly decide to ignore the result.
Nevertheless, this problem can occur in general and users of \name{} should be aware of the possibility of mis-predictions.

\section{Related Work}
\label{sec:related}


We already provided a brief summary of the related work in the introduction to this paper (see Section~\ref{sec:intro} and Table~\ref{table:related}), and the constraints from our setting that forced us to build \name{}. Here we elaborate selectively on some of the related work, starting with the most closely related.

\paragraph{Localizing Bugs to Commits}

Locus addresses the same problem in an IR-based approach that compares the words associated with a commit and the words in a bug report~\cite{Wen2016}.
Orca also takes an IR-based approach and reports deploying a system for a large-scale, distributed, industrial system~\cite{Bhagwan2018}.
Their focus is on handling the frequent re-builds of the system through a build provenance graph.
We compared \name{} directly against prototypes of both Orca and Locus that we built ourselves, and found \name{} to work better for us.
Given that \name{} is building upon IR-based methods, we reason that its primary benefit comes from its ability to handle complex entities and understanding semantics.
This is expected, as Locus and Orca were not designed with these considerations which emerged uniquely in our setting.

ChangeLocator takes a learning-based approach that ranks all commits that change at least one method that appears in a stack trace~\cite{Wu2018}.
In addition to most techniques in Table~\ref{table:related}, ChangeLocator also relies on statically analyzing the code base, which \name{} avoids due to scalability issues.

\paragraph{Localizing Bugs to Files}

A majority of previous work that uses IR for bug localization does so for files~\cite{Nguyen2011,Zhou2012,Saha2013,Wu2014,Ye2014,wong2014boosting,Moreno2014,Lam2015,Rahman2018,Koyuncu2019}.  This body of work has made the important connection that information retrieval techniques can be very fruitful in software engineering, where discovering relationship between different kinds of artifacts arises in many ways, and formal techniques using program dependence have not been as effective. We have built upon this work to adapt it to satisfy the constraint of our industrial setting.  Not only did we need to use bug localization to commits rather than files, we also needed to factor in the high-level structure of the documents in how we modeled the problem.

\paragraph{Clustering and Prioritizing Crashes}

The potentially large number of crashes revealed by fuzz-testing or widely deploying software has motivated work on clustering crashes, e.g., based on similarities of stack traces~\cite{Dhaliwal2011,Dang2012}, on repairs that prevent a crash~\cite{DBLP:conf/kbse/TonderKG18}, or other heuristics~\cite{DBLP:journals/cacm/KinshumannGGAONGLH11}.
Once crashes are clustered, \cite{DBLP:conf/sigsoft/CastelluccioSVP17,qian2020ccsm} propose to help understanding the crash by identifying features that are unique to a cluster.
Another line of work prioritizes crashes, e.g., based on the distribution of occurrences among users~\cite{DBLP:conf/wcre/KhomhCZH11} or based on a prediction of how likely a crash will occur for other users~\cite{DBLP:journals/tse/KimWKZCP11}.
All this work is orthogonal to the problem addressed here, and could be used before localizing the culprit commit for a bug with \name{}.

\paragraph{Coverage-based fault localization}

A very popular line of previous work takes the position that a fine grain fault localization -- down to lines of code -- can be carried out, given a statement-level coverage of test executions, both passing tests and failing tests~\cite{Jones2002, abreu2007accuracy, DBLP:conf/icse/LiZdO18}.  These methods try to recover the correlation of the failing tests to certain lines of code.  This work is a bit orthogonal to our work, in that it may apply when it comes to an engineer trying to debug a particular crash, but it does not give a quick way to identify a blamed commit which must be either debugged or reverted.  
The main difference to coverage-based bug localization is that \name{} does not require any coverage information, but only a crash trace that manifests the bug.  Furthermore, the usefulness of these methods in debugging has lately been questioned by Parnin and Orso~\cite{Parnin2011}.

\paragraph{Other Related Work}

Many techniques for automated program repair~\cite{cacm2019-program-repair} rely on localizing where to fix a bug.
\name{} could serve as a starting point for repair of bugs that manifest through crashes.
\cite{DBLP:journals/ese/JonssonBBSER16} address the problem of assigning a bug report to a developer team.
%
Some of the past work in bug localization uses inputs beyond bug reports or crash traces, e.g., by also considering meta-data from version histories, such as files involved in fixing past bugs~\cite{Ye2014,Lam2015}.
\name{} does not require this kind of meta-data, but could possibly benefit from it.

\section{Conclusion}
\label{sec:conclusion}

We conclude with a summary of our perspective on IR-based bug localization by introspecting what we have accomplished in this work.
\name{} does not solve a new problem, it in fact builds upon existing work in IR-based bug localization.
Instead, what we have done is evaluate these methods at a real world large-scale industrial setting, Facebook, with practical considerations that would not typically arise in other contexts.
Our evaluation has revealed shortcomings of existing approaches.
Utilizing a suite of enhancements, \name{} pushes the boundary of IR-based localization in practice.
We have shown that \name{} is better in two applications, but we have already found one application where more work had to be done to boost its performance higher and make a significant impact to Facebook's developer operations.

In addition to performance, there are also secondary concerns that are orthogonal to this work.
For example, there exists a trade-off between complexity of a model and its interpretability -- the word embeddings model might be good at understanding semantics, but it is not as easily interpretable as simple word weights from BM25 or tf-idf.
Interpretability of a model is important for engineers to build trust in its predictions, independent of how accurate the model can be.
Owing to these considerations, we posit that there is still more room for improvement for industrial applications of IR-based localization methods, and that they are not a panacea, based on our perspective from Facebook.

\printbibliography

\end{document}